\documentclass[preprint,showpacs,preprintnumbers,amsmath,amssymb]{revtex4}
\usepackage{amsmath}
\usepackage{amssymb}
\usepackage{epsf}
\usepackage{color}
\usepackage[dvips]{graphicx}

\usepackage{graphicx}
\usepackage{dcolumn}

\newcommand{\ud}[1]{#1^{\dag}}
\newcommand{\bra}[1]{\left\langle #1\right|}
\newcommand{\ket}[1]{\left| #1\right\rangle}

\newcommand\sign[1]{\mathrm{sign}(#1)}

\begin{document}

\title{Multiplets in Emission of Large Quantum Dots in Microcavities}

\author{Fabrice~P.~Laussy}
\author{Alexey Kavokin}
\author{Guillaume Malpuech}
\affiliation{%
Lasmea, CNRS, Universit\'e Blaise Pascal -- Clermont--Ferrand II\\24 avenue des Landais, 63177 Aubi\`ere Cedex, France}
\date{\today}

\begin{abstract}
  We show theoretically that the emission spectrum of a single large
  quantum dot strongly coupled to a single photon mode in a
  microcavity can be qualitatively different from the spectrum
  obtained with an atom in a cavity.  Instead of the well--known
  Mollow triplet we predict appearance of multiplets with the number
  of peaks a function of the quantum dot size and pumping intensity.
  The mutiplets can appear if the quantum dot is larger than the
  exciton Bohr radius, so that excitons are confined as whole
  particles in the dot. In this case the Pauli principle is relaxed
  and one can accommodate more than one exciton (but still a finite
  number) in a given quantum state.
\end{abstract}

\maketitle

\emph{Introduction} --- Coupling of a zero-dimensional photonic mode
to a zero-dimensional exciton (electron-hole) state has been realized
in pillar and photonic microcavities containing quantum
dots~\cite{gerard98a, solomon00a, reithmaier04a, yoshie04a}. Very
recently, two groups have obtained independently the experimental
evidence for vacuum-field Rabi splitting in such
structures~\cite{reithmaier04a, yoshie04a}.  These observations
confirmed theoretical predictions of strong exciton-photon coupling in
zero-dimensional systems~\cite{andreani99a, kaliteevski01a}. 

Nonlinear optical effects in strongly coupled zero-dimensional systems
are expected to be extremely rich. For a while, only a manifestation
of the Purcell effect~\cite{purcell46a} had been experimentally
reported~\cite{gerard98a, solomon00a}, while theoretical predictions
went much beyond, evoking appearance of dressed exciton and bi-excitons
states~\cite{panzarini99a}.  The crucial question for description of
emission from quantum dot cavities (QDC) is whether crystal
excitations coupled to light behave like fermions or like bosons in
these systems~\cite{glazov}. The fermionic regime has been a subject
of cavity quantum electrodynamics (QED) for decades.
It is familiar in atomic cavities, and can be efficiently described by
Dicke formalism~\cite{Dicke}. Its signature is the appearance of the
Mollow triplet~\cite{mollow69a} in emission spectra of the cavity.

Clearly, the fermionic regime is realized if carriers are strongly
confined in small-size quantum dots (case of~\cite{yoshie04a}, for
example). On the other hand, it does not hold if the dot size is much
larger than the exciton Bohr radius (further referred to as the
``large dot'' case, realized in particular in the dots originated from
islands of quantum well width
fluctuation\cite{reithmaier04a,gammon96a}).  In large dots, excitons
are quantized as whole particles and exhibit bosonic properties at
least to some extent. The limiting case of a large dot is a quantum
well where quasi-Bose condensation of excitons is indeed
possible~\cite{lozovik75a} and Rabi-doublets are observed in the
spectra of microcavities~\cite{kavokin03b}. Dicke formalism does not
hold in this case, as it does not take into account the fundamental
difference between an atomic and an excitonic system: the number of
atoms is fixed while the number of excitons is not.  From a formal
point of view in the purely fermionic case the light mode is coupled
to a two-level system, while in the purely bosonic case it is coupled
to a harmonic oscillator having an infinite number of equidistant
energy levels.  Logically, the straightforward intermediate case would
correspond to the optical mode coupled to a truncated harmonic
oscillator, having a finite number, say~$N$, of energy levels. As such,
a large QD offers a realization of parastatistics which
interpolate between Bose-Eistein and Fermi-Dirac statistics by
allowing up to~$N>1$ particles in a given quantum state~\cite{gentile,
  green52a}.

In this work we study theoretically the emission of light by a QDC
using the model of a truncated harmonic oscillator. We assume that a
finite number of excitons ($N\ge 1$) can be generated in the quantum
state of interest. Here $N$ is taken as a free parameter, expected to
scale like~$S_\mathrm{dot}/\pi a^2_\mathrm{B}$ where~$S_\mathrm{dot}$
is the cross-section surface of the dot and $a_\mathrm{B}$ the exciton
Bohr-radius. Physically, it corresponds to the following image: the
dot is much larger than the exciton Bohr radius, so that excitons are
confined as whole particles inside it. $N=1$ recovers the fermionic
case. We also assume that the energy separation between quantum
confined excitonic states exceeds the vacuum Rabi splitting, so that
one can neglect all excitonic states except one, strongly coupled to a
single light mode.  This latter condition, though very artificial for
quantum well microcavities, is easily met in QDC, where typical values
of the Rabi splitting are tens of $\mu$eV~\cite{reithmaier04a,
  yoshie04a}.  Furthermore, we neglect exciton-exciton interactions,
which allows us to assume the energy of the excitonic transition to be
independent on the number~$m$ of excitons already created in a given
quantum state if~$m<N$. If~$m=N$ no more excitons can be created in
the quantum state we consider.  Finally, we assume the exciton-photon
system to be fully spin-polarized, so that polarization effects can be
neglected and bi-excitons bound state impossible. Experimentally this
corresponds to the resonant circularly polarised pumping and no
efficient spin-relaxation time in the system.
In general to deal with QDC one should solve the complicated
many-particles problem accounting for exciton-exciton, exciton-carrier
and exciton-photon interactions. Here we deal with the simplified
model described above as an ideal limit of more refined microscopic
calculations~\cite{glazov} which display quantitative departures but
an overall qualitative agreement.  Our present model retains the
essential physical difference of a large quantum dot from a small one
and a quantum well and allows to obtain analytically the emission
spectrum of the structure. We show that this spectrum is remarkably
different from well-known Mollow-triplet and Rabi-doublet.  It
exhibits multiplets whose number is sensitive to the dot size and
pumping intensity. 

\emph{Formalism} --- The electromagnetic field is modelled in the
approximation of the quasi--mode coupling by single mode Bose field
annihilation operator~$a$. We introduce for a large quantum dot, able
to host~$N$ excitons, the operator~$\sigma_N$ which behaves like a
Bose field annihilation operator when the number of excitations is
less than or equal to~$N$, i.e.,
\begin{equation*}
  \sigma_N\ket{m}=\sqrt{m}\ket{m-1},\quad\ud{\sigma_N}\ket{m}=\sqrt{m+1}\ket{m+1}\theta_{m,N}
\end{equation*}
where~$\theta_{m,N}$ is the modified Heaviside step function which is
zero if~$m\ge N$ and one if~$m<N$.  Its matrix representation in the
basis of Fock states is the $N+1$ squared matrix which is a truncation
of~$\sigma_{\infty}$ the annihilation matrix of a Bose field. For
instance, for~$N=2$,
\begin{equation}
  \sigma_2=
  \begin{pmatrix}
    0 & 1 & 0 \\
    0 & 0 & \sqrt2 \\
    0 & 0 & 0
  \end{pmatrix}\,. 
\end{equation}
The algebra for this operator follows straightforwardly:
\begin{equation}
  \label{eq:commutation}
  [\sigma_N,\ud{\sigma_N}]=\mathbf{1}_{N+1}-\lambda_{N+1}\,,
\end{equation}
where~$\mathbf{1}_{N+1}$ is the~$N+1$ squared identity matrix
and~$(\lambda_{N+1})_{i,j}=(N+1)\delta_{i,j}\delta_{i,N}$ (with~$0\le
i,j\le N$). 

The hamiltonian for our system is similar to Dicke~\cite{Dicke} or
Jaynes--Cummings~\cite{jaynes63a} hamiltonian but instead of coupling
the field~$a$ with a spinor~$J$ or a two--level system
(fermion)~$\sigma_1$, one couples~$a$ with this operator~$\sigma_N$,
also in the rotating wave approximation:
\begin{subequations}
  \begin{eqnarray}
    H&=&\hbar\omega\ud{a}a+\hbar\omega\ud{\sigma_N}\sigma_N\label{eq:H1}\\
    &+&\hbar g(\ud{a}\sigma_N+a\ud{\sigma_N})\,.\label{eq:H2}
  \end{eqnarray}
\end{subequations}
We considered zero detuning for simplicity. The general case with a
small detuning brings fine quantitative departures which are the topic
for a separate communication. Here~$\hbar g$ is the coupling
constant corresponding to (half) the Rabi splitting in the linear
regime~\cite{reithmaier04a, yoshie04a}. First line~(\ref{eq:H1})
describes bare states of photons and excitons, the later being limited
to a maximum $N$. We note eigenstates of this first line
alone~$\ket{n,m}$ with~$n$ the number of photons, any integer number,
and~$m$ the number of excitons, any integer less than or equal to~$N$. 
Second line~(\ref{eq:H2}) describes the coupling between the two
fields which in absence of dissipation will always be ``strong'' in
the quantum mechanical sense.  The dissipation is caused by two
channels of energy escape included as non-hermitian contributions of
the Lindblad type to Von Neumann equation,
\begin{equation}
  \label{eq:liouville}
  i\hbar\dot\rho=[H,\rho]+\mathcal{L}_a\rho+\mathcal{L}_{\sigma_N}\rho\,,
\end{equation}
where
\begin{equation}
  \label{eq:lindblad}
  \mathcal{L}_A\equiv-{\gamma_A\over2}(\ud{A}A\rho-2A\rho\ud{A}+\rho\ud{A}A)
\end{equation}
for~$A=a$ (resp.~$\sigma_N$) describing the photon leakage through the
cavity mirror by tunnel effect (resp.~the emission from the active
state within the cavity) with associated decay constant~$\gamma_a$
(resp.~$\gamma_{\sigma_N}$).  The procedure is straightforward in
principle: since the number of excitations is conserved by~$H$, one
diagonalises it in the subspace of~$n$ excitations and thus obtains
new eigenstates of the system (so--called \emph{dressed states} in
cavity QED terminology).  Thus, in the
basis~$\mathcal{H}_n=\{\ket{0,n},\ket{1,n-1},\ldots,\ket{N,n-N}\}$
(for~$n\ge N$), the hamiltonian reduces to:
\begin{eqnarray}
  \label{eq:generalhamiltonian}
  H&=&\hbar\omega n\mathbf{1}_{N+1}\notag\\
  &+&\hbar g \Big[\sqrt{i(n-i+1)}\delta_{i,j+1}\\
  &+&\phantom{\hbar g\Big[}\sqrt{(i+1)(n-i)}\delta_{i,j-1}\Big]_{0\le i,j\le N}\notag
\end{eqnarray}
where the term between square brackets is the generic expression for
the interaction matrix (which is zero everywhere but below and above
its diagonal). Case~$N=1$ gives the Mollow triplet. 

Along with general results, we will deal explicitely with the
all--important case~$N=2$. It is expected the effect we predict will
first be observed for this very value~$N=2$. In this case, the matrix
representation~(\ref{eq:generalhamiltonian}) in the
basis~$\mathcal{H}_2$ reads for~$n$ excitations (with~$n\ge 2$)
\begin{equation}
  H=\hbar\omega n\mathbf{1}_3+\hbar g
  \begin{pmatrix}
    0 & \sqrt n & 0 \\
    \sqrt n & 0 & \sqrt{2(n-1)} \\
    0 & \sqrt{2(n-1)} & 0 \\
  \end{pmatrix}\,,
\end{equation}
Cases where~$n$ is lower than or equal to~$N$ fall into linear regime
displaying two peaks, the exciton field never reaching its saturation
density.  If~$n>N$, the new~$N+1$ eigenstates (dressed states) will be denoted,
for the manifold~$n$, by~$\ket{\nu}_n$ with~$\nu=1,\dots,N+1$ indexing
the state. For~$N=2$,
\begin{subequations}
  \label{eq:coeffs}
  \begin{eqnarray}
  \ket{\nu=1}_n&\equiv&\left(\sqrt{n\over2(3n-2)},-{\sqrt 2\over2},\sqrt{n-1\over3n-2}\right),\\
  \ket{\nu=2}_n&\equiv&\left(-\sqrt{2(n-1)\over3n-2},0,\sqrt{n\over3n-2}\right),\\
  \ket{\nu=3}_n&\equiv&\left(\sqrt{n\over2(3n-2)},{\sqrt 2\over2},\sqrt{n-1\over3n-2}\right). 
  \end{eqnarray}
\end{subequations}
in the bare states basis~$\mathcal{H}_2$. The associated eigenvalues are:
\begin{equation}
\label{eq:Ennu}
E^n_\nu\equiv\hbar\omega n+(\nu-2)\hbar g\sqrt{3n-2}  
\end{equation}
The spectrum of emission comes from transitions between manifolds
with~$n$ and~$n-1$ excitations. One can obtain the exact expression
for the luminescence spectrum from~(\ref{eq:liouville}) and quantum
regression theorem to evaluate the time correlation function needed
for Wiener-Khintchin theorem. We content here with Fermi's golden rule
formula for perturbations responsible for~(\ref{eq:lindblad}). These
perturbations arise from the weak coupling of cavity photons~$a$
(resp.~excitons $\sigma_N$) with the continuum of external modes of
the electromagnetic field in vacuum state.  Therefore,
calling~$c_i^{\nu,n}$ the projection of state~$\ket{\alpha}_n$ on
state~$\ket{i,n-i}$ one can obtain the coefficients
in~(\ref{eq:coeffs}) as:
\begin{equation}
  \ket{\nu}_n=\sum_{i=0}^N c_i^{\nu,n}\ket{i,n-i} 
\end{equation}
The corresponding transition probabilities between
states~$\ket{\nu,n}$ and~$\ket{\nu',n-1}$ are obtained as:
\begin{eqnarray}
  \label{eq:strength}
    I_\mathrm{end}&=&|\bra{\nu'}_{n-1}a\ket{\nu}_n|^2=\left|\sum_{i=0}^N(c_i^{\nu',n-1})^* c_i^{\nu,n}\sqrt{n-i}\right|^2\label{eq:reflectivity}\notag\\
    I_\mathrm{lat}&=&|\bra{\nu'}_{n-1}\sigma_N\ket{\nu}_n|^2=\left|\sum_{i=1}^N(c_{i-1}^{\nu',n-1})^* c_i^{\nu,n}\sqrt{i}\right|^2
\end{eqnarray}
In cavity QED terminology, $I_\mathrm{end}$ and~$I_\mathrm{lat}$
correspond to \emph{end--emission} and~\emph{lateral--emission}
photo--detection respectively, while in luminescence of a microcavity,
one observes the sum of the two
contributions~$I_\mathrm{lat}+I_\mathrm{end}$ simultaneously. To
detect~$I_\mathrm{lat}$ independently one should use the scattering
geometry. 

\begin{figure}
    \epsfxsize=9cm\epsfbox{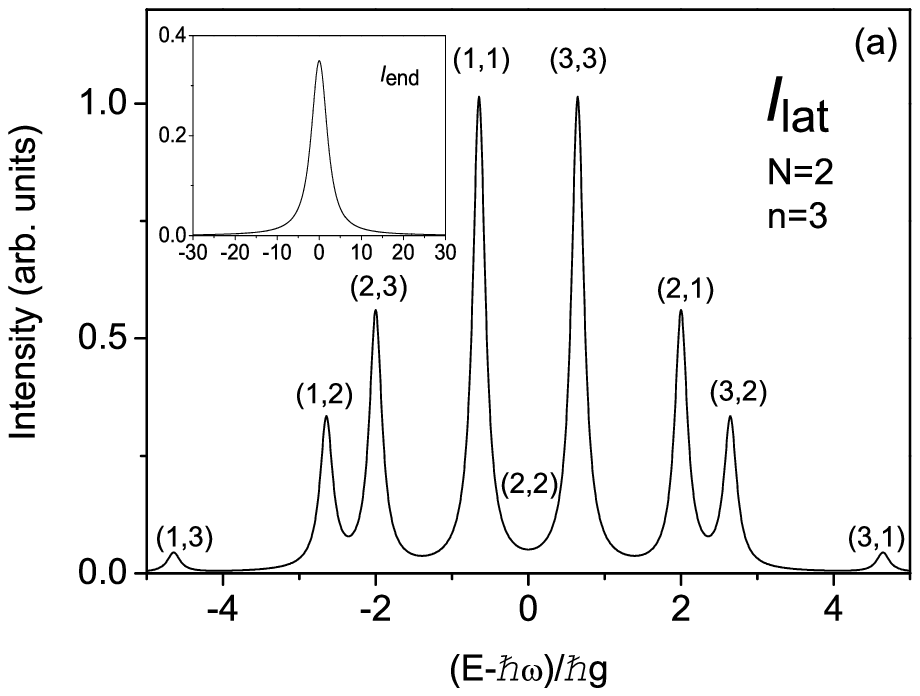}
    \epsfxsize=9cm\epsfbox{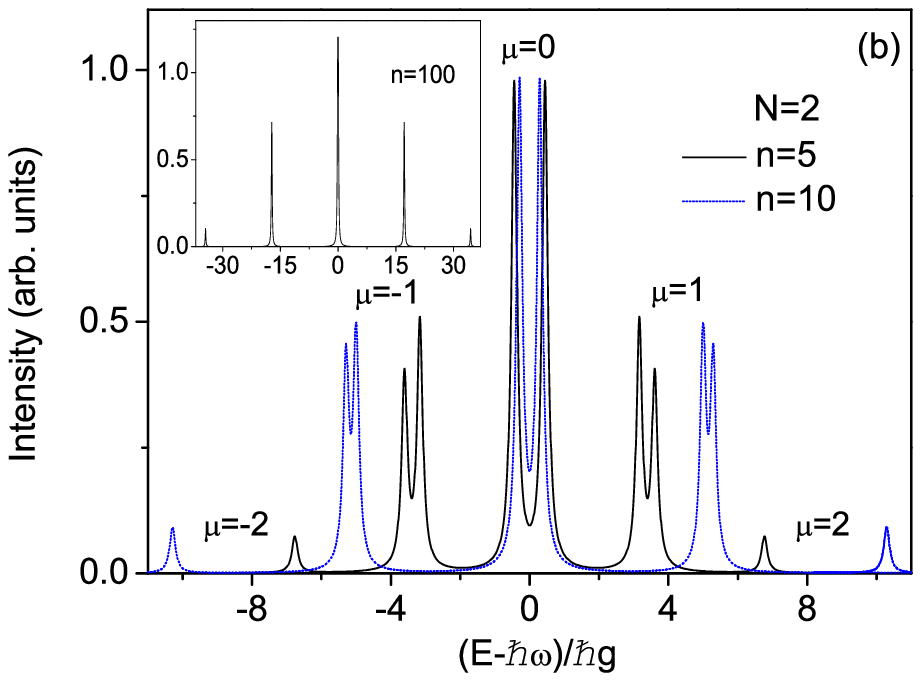}    
    \caption{Multiplets observed in the spectrum~$I_\mathrm{lat}$ of
      lateral emission of the cavity for~(a) $N=2$ and~$n=3$ and~(b)
      $n=5$ (solid) and $n=10$ (dashed) (also for~$N=2$).  The
      coupling constant~$\hbar g=50\mu$eV.  Broadening is obtained by
      convolution with Lorentz functions of width~$\gamma_a=100\mu$eV
      and~$\gamma_{\sigma_2}=5\mu$eV~\cite{reithmaier04a}. The inset
      in Fig.~(a) showns~$I_\mathrm{end}$ for~$n=3$, which is a
      Lorentz function; the inset of Fig.~(b) shows~$I_\mathrm{lat}$
      in the limit of huge~$n$ ($n=100$).} 
  \label{fig:fig1}
\end{figure}

\emph{Results} --- Our main result is the appearance of multiplets in
the configuration~$n\ge N$. This is in stark contrast with cavity QED
where nonlinearity results in apparition of the \emph{Mollow
  triplet}~\cite{mollow69a}). Three questions need be solved in the
general case: first the number of peaks allowed by the transitions
which is of a combinatorial nature, providing the number of
transitions between two multiplets taking into account their
degeneracy. Second the energy splitting between the various peaks thus
obtained which is the eigenvalues problem. Third the strength of
transitions which is the eigenvectors problem. If the radiative
broadening is so small that all peaks can be discriminated and if no
selection rules forbid some transitions, the multiplets structure is
that of~$2N+1$ groups of more or less closely packed peaks for a total
number of~$(N+1)^2$. We label~$\mu=0$ the group at the centre of the
spectrum, and~$\mu=\pm1, \pm2,\dots$ groups located symmetrically
about the central one. The number of peaks is~$N+1-|\mu|$ in
the~$\mu^\mathrm{th}$ group. Each peak is identified unambiguously by
a couple of integers~$(k,l)$ with~$0\le k,l\le N$.  Peak $(k,l)$
belongs to group~$\mu=k-l$ and is the~$k^\mathrm{th}$
(resp.~$l^\mathrm{th}$) if located at the left (resp.~right) of the
spectrum, counting as first in each group that of lower energy. The
energy of this peak~$E(k,l)$ is given by the photon energy released in
the transition between two manifolds and is therefore given by the
difference of initial and final eigenvalues~$E_k^n$ in the dressed
states basis:
\begin{equation}
  \label{eq:ekl}
  E(k,l)=E_k^n-E_l^{n-1}
\end{equation}
In the case~$N=2$, from~(\ref{eq:Ennu}) and~(\ref{eq:ekl}) we get in good
approximation (the better the higher~$n$) the mean energy of
peak~$(k,l)$ as
\begin{equation}
  \label{eq:nine}
  E(k,l)\approx\hbar\omega+(k-l)\hbar g\sqrt{3n}-(2k-5l+6){\hbar g\over2\sqrt{3n}}
\end{equation}
This gives the structure of the multiplet: there is a
splitting of magnitude~$|\nu|\hbar g\sqrt{3n}$ between the central
($\mu=0$) and the $\nu$th group of peaks. The central group has a fine
structure splitting (between (1,1) and~(3,3)) of~$\sqrt 3\hbar
g/\sqrt{n}$. At the onset of the nonlinear effect ($n=3$) the splitting
is close to the Rabi splitting~$2\sqrt{(\hbar
  g)^2-(\gamma_a-\gamma_{\sigma_2})^2/16}$.  The fine structure
splitting of the groups~$\mu=\pm2$ is half the central one: $\sqrt
3\hbar g/(2\sqrt n)$. As~$n$ increases, each group of peaks gets
farther from the others while inside this group, the fine structure
splitting is reduced. This behaviour is seen in Fig.~(\ref{fig:fig1}). 
When the photon field is so intense that for the given
broadening~$\gamma_{\sigma_2}$ the fine splitting is unobservable, the
spectra resemble a Mollow triplet. It can be distinguished from it by
the splitting which is~$\sqrt 3/2$ times smaller and the presence of
additional (small) peaks with a splitting~$\sqrt{3}$ larger. Also the
ratio between the central peak and its closest satellites is~$9/4$
rather than~$1/2$ in case of the Mollow triplet (see
Fig.~(\ref{fig:fig1}b)). 

This structure applies both for~$I_\mathrm{end}$ and~$I_\mathrm{lat}$
spectra.  However, they display two drastically different behaviours. 
First $I_\mathrm{end}$ is maximum at the center of the spectrum with
peak~(2,2) the highest one. The situation is opposite
with~$I_\mathrm{lat}$ for which, whatever~$n$, peak~(2,2) is an
exactly forbidden transition. 

\begin{figure}
    \epsfxsize=9cm\epsfbox{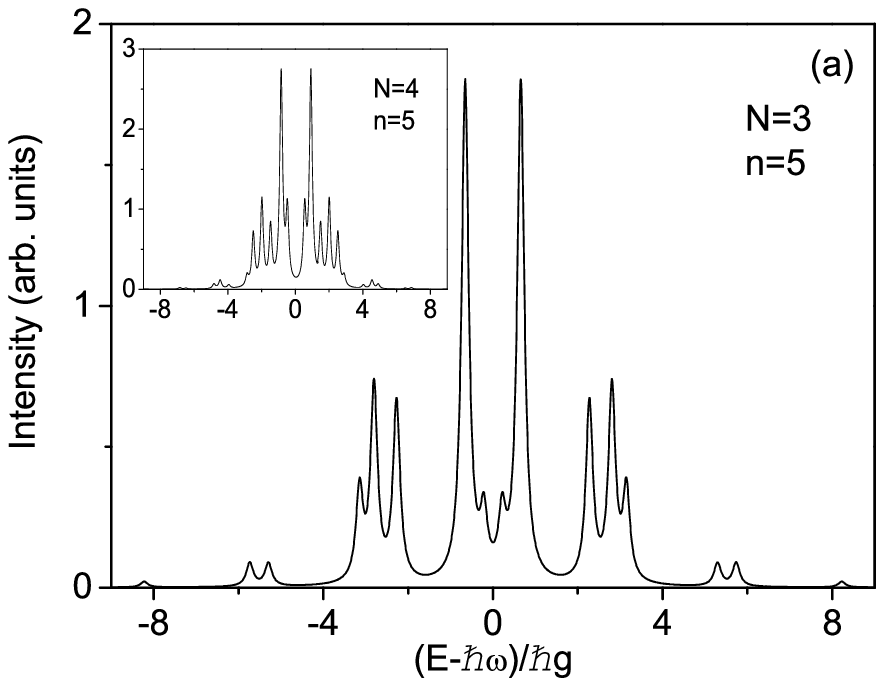}
    \epsfxsize=9cm\epsfbox{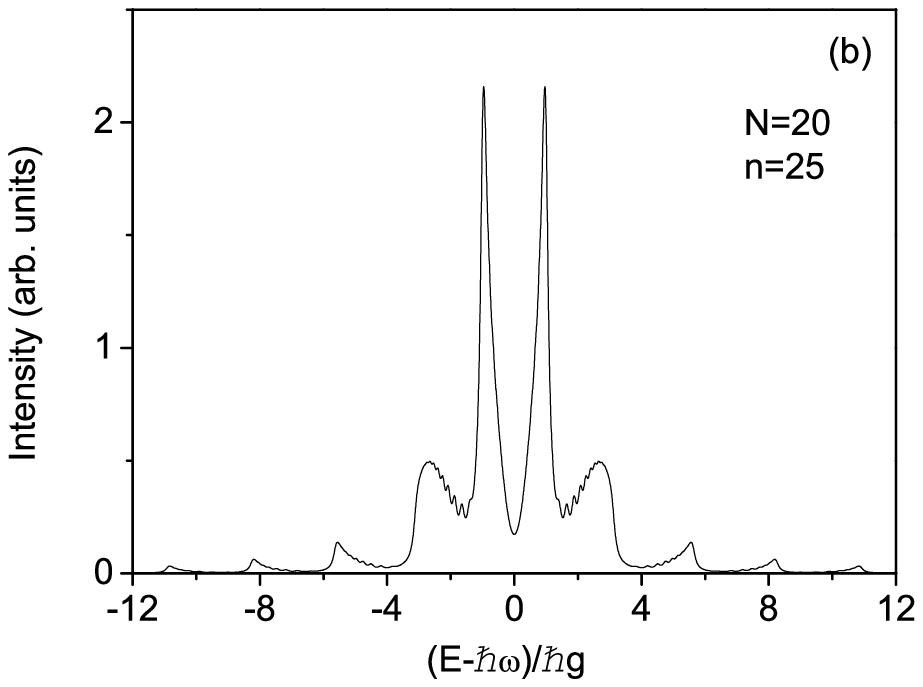}
    \caption{Multiplets for~(a) $N=3$, $n=5$ (inset $N=4$, $n=5$)
      and~(b) $N=20$, $n=25$ with same parameters as
      in~Fig.~(\ref{fig:fig1}).} 
  \label{fig:fig2}
\end{figure}

The same procedure can be applied to larger values of~$N$, e.g.,
for~$N=3$ (cf.~\ref{eq:Ennu}):
\begin{equation}
  E_\nu^n=\hbar\omega n+\sign{\nu-{5\over2}}\hbar g\sqrt{(-1)^\nu\sqrt 2\sqrt{3n^2-9n+8}+3n+4}\label{eq:N3}  
\end{equation}
with~$1\le\nu\le4$, and for~$N=4$:
\begin{equation}
  \begin{split}
    E_\nu^n&=\hbar n\omega+\sign{\nu-3}\hbar g\times\\&\sqrt{\sign{3-\nu}(-1)^{\nu}\sqrt{2}\sqrt{5n^2-25n+38}+5n-10}\,,\label{eq:N4}
  \end{split}
\end{equation}
this time with~$1\le\nu\le5$, etc\dots\ (where $\sign x=0$ if $x=0$
and is~$x/|x|$ otherwise.) Along the same lines as before, one can
derive the structure and splittings of the spectrum of emission in
these cases. For instance, for~$N=3$, the (1,1)--(3,3) splitting
is~$(\sqrt{3-\sqrt6})\hbar g/\sqrt n\approx0.74\hbar g/\sqrt n$. Note
that the difference in energy enters as square roots inside a square
root and thus becomes smaller and smaller with increasing~$N$. 
Figure~(\ref{fig:fig2}b) shows the case of an extremely large dot
where satellite peaks become suppressed and the Rabi-like doublet
dominates the spectrum.  For~$N\rightarrow\infty$ the familiar
Rabi-doublet seen in the spectra of quantum well cavities is recovered
by this model. 

\emph{Conclusions} --- We predicted appearance of multiplets in
emission spectra of zero-dimensional photonic cavities containing
single large quantum dots. The effect comes from relaxation of the
Pauli principle for excitons confined as whole particles in large
dots.  The effect is strictly nonlinear as it requires the number of
photons injected in the cavity to exceed the capacity of the dot. This
capacity governs the number of peaks in emission. In the small dot
limit, it reduces to three (Mollow triplets). 

\emph{Acknowledgements} --- We thank Dr.~Ivan Shelykh for his thorough
discussion of our formalism. This work has been supported by the
Clermont-2 project MRTN-CT-2003-S03677. 

\bibliography{laussy}

\end{document}